\documentclass{ws-ijmpcs}
\begin{document}

\markboth{Zilong Chang}
{Gluon Polarization at STAR}

%
\catchline{}{}{}{}{}
%

\title{Gluon Polarization in Longitudinally Polarized $pp$ Collisions at STAR
}

\author{Zilong Chang for the STAR Collaboration
}

\address{Cyclotron Institute, Texas A\&M University\\
College Station, Texas 77843-3366,
USA
\\
changzl@physics.tamu.edu}

\maketitle

\begin{history}
\received{Day Month Year}
\revised{Day Month Year}
\published{Day Month Year}
\end{history}

\begin{abstract}
The STAR Collaboration is performing a wide range of measurements to determine the
gluon helicity distribution in the proton.  Gluon-gluon and quark-gluon scattering dominate jet production in proton-proton collisions at RHIC, so the longitudinal double-spin asymmetry, $A_{LL}$, for jet production places significant constraints on the gluon polarization in the proton. In recent years STAR has recorded large longitudinally polarized $pp$ data sets at both $\sqrt{s} = 200$ GeV and $510$ GeV. The 2009 STAR inclusive jet $A_{LL}$ measurements at $\sqrt{s} = 200$ GeV show the first experimental evidence of non-zero gluon polarization over the Bjorken-$x$ range, $x > 0.05$. Furthermore, data collected at $\sqrt{s} = 510$ GeV during the 2012 and 2013 RHIC runs allow access to the gluon polarization at lower $x$. In this talk, I will present the final results of the 2009 inclusive jet $A_{LL}$ measurement at $200$ GeV, the analysis status of the 2012 inclusive jet $A_{LL}$ measurement at $510$ GeV, and the status of di-jet and other STAR measurements that are sensitive to gluon polarization.
\end{abstract}


\section{Introduction}	
Gluon polarization, \(\Delta G\), is a fundamental and key ingredient to the proton spin sum rule, \(\frac{1}{2} = \frac{1}{2} \Delta \Sigma + \Delta G+ L_{q} + L_{g}\).\cite{spinsum} Polarized deep inelastic scattering (DIS) has measured quark and anti-quark polarization, \(\Delta \Sigma \approx0.3\),\cite{dssv}\cdash\cite{dssv06} but provides only weak indirect limits on \(\Delta G\). The Relativistic Heavy Ion Collider (RHIC) is capable of constraining \(\Delta G\) over the range \(x>0.01\), where \(x\) is the gluon momentum fraction in the proton. The Solenoid Tracker at RHIC (STAR) is performing a wide range of measurements to constrain gluon polarization, including inclusive jets, inclusive \(\pi^{0}\) and di-jets.
 
The longitudinal double spin asymmetry, \(A_{LL}\), for jets and \(\pi^{0}\) is defined as the ratio of the difference of differential cross sections when proton beams have the same and opposite helicities over the sum of the two cross sections.  Jet and \(\pi^0\) production in \(pp\) collisions arise from three parton scattering processes: gluon-gluon (\(gg\)), quark-gluon (\(qg\)) and quark-quark (\(qq\)). Next-to-leading-order (NLO) perturbative Quantum Chromodynamics (pQCD) indicates that in RHIC kinematics \(qg\) and \(gg\) dominate. The partonic longitudinal double spin asymmetry is large in this kinematic range.\cite{jetsub1}\cdash\cite{jetsub2} Therefore jet and \(\pi^{0}\) \(A_{LL}\) are sensitive to the polarized gluon distribution in the proton, making STAR ideal to explore gluon polarization.

RHIC is a world-leading facility capable of colliding either longitudinally or transversely polarized proton beams at energies up to \(\sqrt{s}=510\) GeV. The major detectors at STAR for jet and \(\pi^{0}\) measurements are the Time Projection Chamber (TPC), Barrel Electromagnetic Calorimeter (BEMC), Endcap Electromagnetic Calorimeter (EEMC), and Forward Meson Spectrometer (FMS).\cite{stardet} The TPC provides high-resolution tracking for charged particles in the magnetic field of 0.5 T with acceptance \(|\eta|<1.3\) and \(2\pi\) in azimuth. The BEMC, EEMC and FMS detect neutral energy over \(2\pi\) in azimuth but covering different \(\eta\) ranges: \(-1<\eta<1\), \(1<\eta<2\) and \(2.5<\eta<4\), respectively. There are several other detectors, including the Vertex Position Detector (VPD), Zero Degree Calorimeter(ZDC) and Beam Beam Counter (BBC), for relative luminosity and local polarimetry measurements.\cite{stardet}

\section{Inclusive Jets Measurements}

STAR has measured the inclusive jet cross section and \(A_{LL}\) in 200 GeV \(pp\) collisions and performed detailed comparisons between the data and Monte Carlo simulations.\cite{run3+4}\cdash\cite{run6} Jet measurements probe gluon polarization at \(x\) as low as 0.05 at \(\sqrt{s}\) = 200 GeV and even lower \(x\) at \(\sqrt{s}\) = 510 GeV. In real data, jets are constructed from detector responses. Pythia\cite{simu} and Geant\cite{geant} simulations are used to generate Monte Carlo events and simulate detector responses to estimate systematic uncertainties. Not only are detector level jets reconstructed from simulated detector responses, but also particle level jets from hadronization and parton level jets from scattered partons are studied from Pythia records to compare with detector level jets.

Previous STAR measurements used the mid-point code algorithm\cite{cone} to reconstruct jets with seed energy 0.5 GeV, and jet cone radius 0.7 in \(\eta - \phi\) space. The inclusive jet \(A_{LL}\) from 2006 data\cite{run6} excludes a number of theoretical scenarios that predict large gluon polarization in the accessible \(x\) range. The DSSV model, the first global analysis with STAR 2006 inclusive jet \(A_{LL}\) included, found a relatively small gluon contribution to the proton spin at the \(x\) region where 2006 STAR data reside, \(0.05 < x < 0.2\).\cite{dssv06}

\begin{figure}[ht]
\centerline{\includegraphics[width=0.6\textwidth]{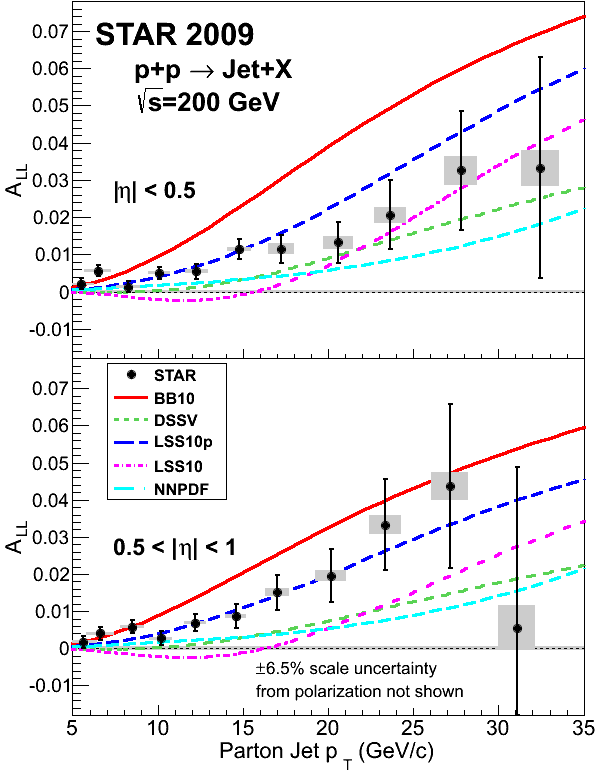}}
\vspace*{8pt}
\caption{STAR 2009 inclusive jet \(A_{LL}\) vs. \(p_{T}\) in two \(\eta\) ranges, \(0<|\eta|< 0.5\) and \(0.5<|\eta|< 1.0\), compared to several recent global analysis predictions.\protect\cite{run9} \label{f1}}
\end{figure}

In 2009, STAR collected nearly a 20 fold increase in event statistics of 200 GeV \(pp\) data compared to the 2006 200 GeV data.  In addition, the anti-\(k_{T}\) algorithm with jet resolution parameter R = 0.6 was used to reconstruct jets. The anti-\(k_{T}\) algorithm is less susceptible to diffuse soft backgrounds from underlying event and pile-up effects compared to the mid-point cone algorithm.\cite{antikt} A change in the treatment of hadronic energy depositions in the BEMC and EEMC led to an improvement in the jet momentum resolution from 23\% to 18\%. The statistical error bars of the 2009 results are at least a factor of four smaller than those of the 2006 results at low jet \(p_{T}\) and a factor of three smaller at high jet \(p_{T}\). As shown in Fig. ~\ref{f1},\cite{run9} the \(A_{LL}\) results fall among several recent model predictions.\cite{bb}\cdash\cite{dssv06} The \(A_{LL}\) measurements fall above the old DSSV prediction but well within the previous quoted uncertainty. The recently released DSSV'14 results with the 2009 data give \(\Delta G = 0.19^{+0.06}_{-0.05} \) for \(x > 0.05\) at 90\% confidence limit.\cite{dssv} The NNPDF group has also included the 2009 results into their fits. They find \(\Delta G\) = \(0.23 \pm 0.07\)  at \(0.05<x < 0.5\) together with a significantly reduced uncertainty band on \( x \Delta g(x)\).\cite{nnpdf}

\begin{figure}[ht]
	\makebox[\textwidth][c]{
	\begin{minipage}{0.6\textwidth}
	\includegraphics[width=\textwidth]{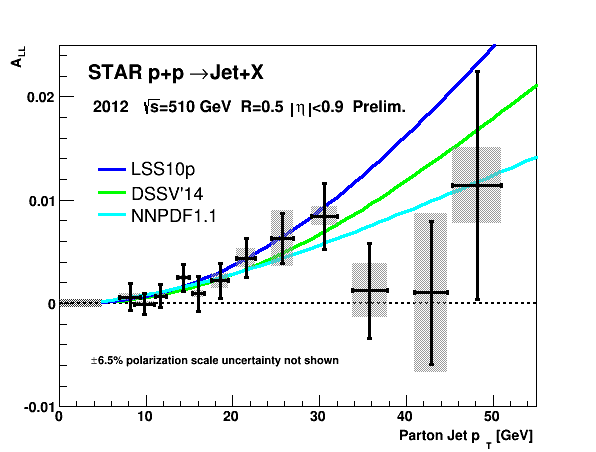}
	\end{minipage}
	\begin{minipage}{0.6\textwidth}
	\includegraphics*[width=\textwidth]{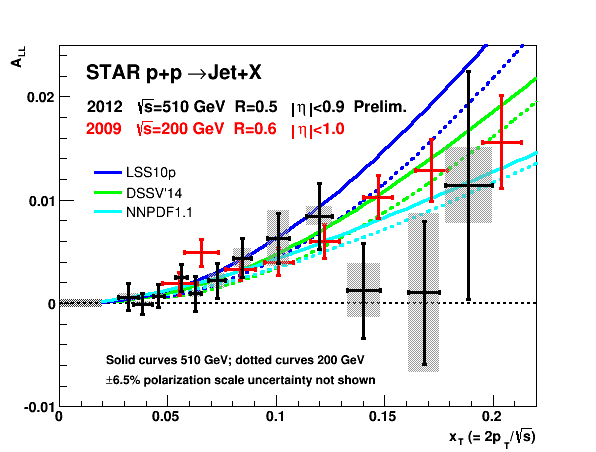}
	\end{minipage}
	}
	\vspace*{8pt}
	\caption{Left: STAR 2012 510 GeV inclusive jet \(A_{LL}\) vs. jet \(p_{T}\) in \(|\eta|< 0.9\). Right: STAR 2012 inclusive jet \(A_{LL}\)and 2009 inclusive jet \(A_{LL}\) vs. \(x_{T}\). \label{f2}}
\end{figure}

In 2012, STAR performed the first measurements of inclusive jet \(A_{LL}\) in 510 GeV \(pp\) collisions. The anti-\(k_{T}\) algorithm with jet resolution parameter R = 0.5 was used to reconstruct jets. The smaller jet resolution parameter 0.5 features reduced backgrounds from underlying event and pile-up effects and better matching probabilities from detector jets to parton jets. The parton jet \(p_{T}\) uncertainty is dominated by the total BEMC energy uncertainty. The dominant systematics in the 2012 \(A_{LL}\) arise from trigger and reconstruction biases, which were estimated with the Monte Carlo simulations. The systematics from non-collision background and residual transverse polarization are found to be negligible. The relative luminosity uncertainty is estimated as 4 \(\times 10^{-4}\). The left panel of Fig. ~\ref{f2} shows the preliminary STAR 2012 510 GeV inclusive jets \(A_{LL}\) results with several of the latest polarized parton distribution function (PDF) model predictions. In the right panel of Fig. ~\ref{f2}, the 2012 510 GeV data are plotted with the 2009 200 GeV data as a function of \(x_{T} = 2p_{T}/\sqrt{s}\). The two sets of data are consistent with each other in the overlapping \(x_{T}\) region. The higher \(\sqrt{s}\) extends gluon helicity measurements to lower \(x\).

\section{Inclusive \(\pi^{0}\) Measurements}

\begin{figure}[ht]
\centerline{\includegraphics[width=0.7\textwidth]{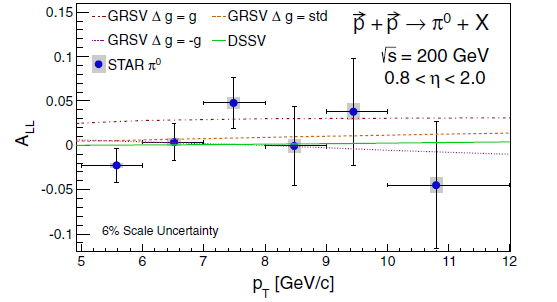}}
\vspace*{8pt}
\caption{STAR 2006 200 GeV inclusive \(\pi^{0}\) at \(0.8<\eta<2.0\).\protect\cite{run6pi} \label{f4}}
\end{figure}

STAR is also performing \(\pi^{0}\) measurements to explore gluon polarization in the proton. It is convenient to reconstruct \(\pi^{0}\) by measuring \(\gamma\) from \(\pi^{0}\) decay. The \(\gamma\)s are detected by the BEMC, EEMC and FMS over a wide \(\eta\) coverage. STAR has measured the inclusive \(\pi^{0}\) cross sections over several \(\eta\) ranges at 200 GeV, such as \(0.0<\eta<1.0\), \(0.8<\eta<2.0\), \(<\eta>=3.3\) and \(<\eta>=3.68\).\cite{midpi}\cdash\cite{fwdpi2} The STAR 200 GeV inclusive \(\pi^{0}\) \(A_{LL}\) at \(0.8 < \eta < 2.0\) probes the gluon helicity density down to \(x=0.02\). Fig. ~\ref{f4} shows the 2006 inclusive \(\pi^{0}\) results.\cite{run6pi} The statistical precision of the 2006 data is not sufficient to discriminate among different models for \(\Delta G\). The 510 GeV data recorded during 2012 will achieve significantly greater precision for inclusive \(\pi^{0}\) \(A_{LL}\). The projected statistical uncertainty is less than 0.015 across the entire \(\pi^{0}\) \(p_{T}\) range. The higher beam energy also will extend the sensitivity to the lower \(x\) gluon helicity density. New measurements at further forward pseudo-rapidity with the FMS are discussed in another article of this journal.\cite{fmspi}

\section{Di-jets Measurements}
\begin{figure}[ht]
	\makebox[\textwidth]{
	\begin{minipage}{0.5\textwidth}
	\centerline{\includegraphics[width=\textwidth]{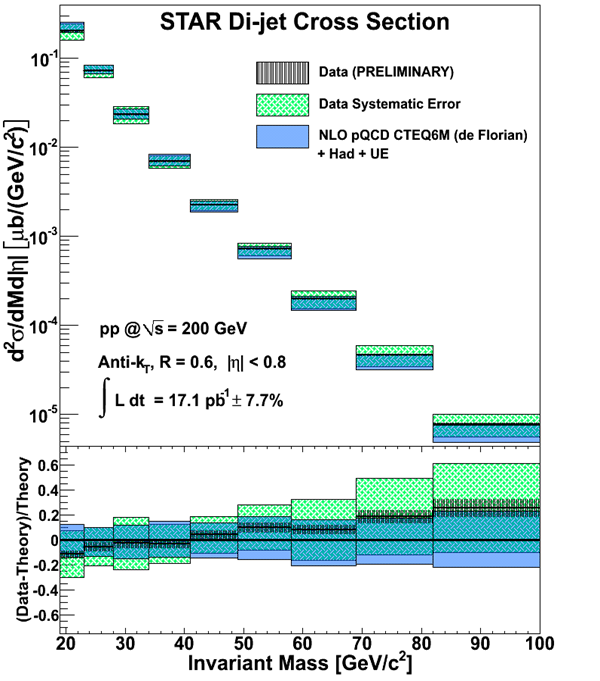}}
	\end{minipage}
	\begin{minipage}{0.5\textwidth}
	\centerline{\includegraphics*[width=\textwidth]{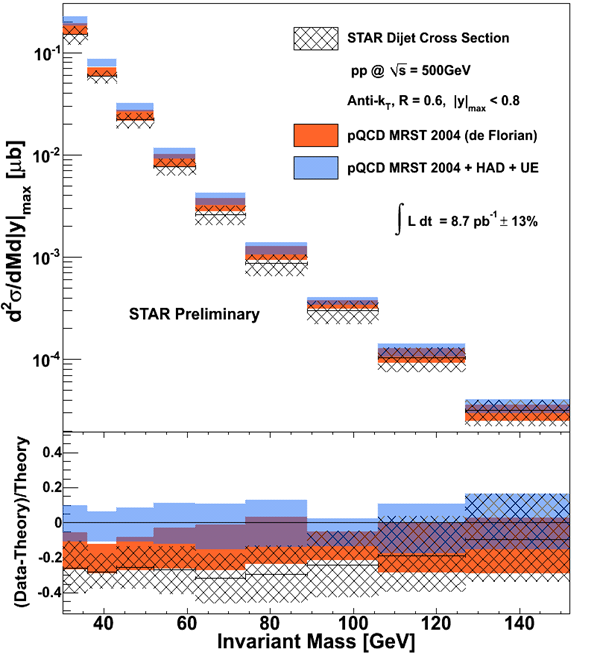}}
	\end{minipage}
	}
	\vspace*{8pt}
	\caption{STAR 2009 200 GeV (left) and 2009 500 GeV (right)\protect\cite{dijet500} di-jet cross sections at \(|\eta| < 0.8\). \label{f5}}
\end{figure}

Di-jet measurements at STAR permit event-by-event calculations of \(x_{1}\) and \(x_{2}\) at leading order. The di-jet cross sections at 200 and 500 GeV are shown in Fig. ~\ref{f5}.\cite{dijet500} Both results agree well with NLO pQCD calculations with corrections for hadronization and underlying event. The 2009 200 GeV dijet \(A_{LL}\) will become available very soon. The 510 GeV di-jet \(A_{LL}\) with large statistics from the 2012 and 2013 data is being analyzed now.

\section{Future and Conclusion}

STAR plans to triple the statistics of the existing 200 GeV data after the 2015 RHIC run. It is anticipated that the new results will significantly reduce the uncertainty for 200 GeV \(A_{LL}\) relative to the 2009 200 GeV results, a factor of about two at jet \(p_{T} > 15\) GeV. STAR also plans to install a Forward Calorimeter System (FCS) around 2020. The FCS will enable to measure one or both of di-jets in the forward region \(2.8<\eta<3.7\). At \(\sqrt{s}=500\) GeV, with one jet in EEMC and one jet in FCS, \(A_{LL}\) measurements will access the gluon helicity density down to \(x \approx 5 \times 10^{-3}\). With both jets in FCS, they access the gluon helicity density at \(x\) below \(10^{-3}\).  Detailed simulations of the expected sensitivities can be found in Ref.~\refcite{fcs}


STAR has developed mature techniques to reconstruct jets and \(\pi^{0}\). The 2009 inclusive jet \(A_{LL}\) results provide the first experimental evidence for positive gluon polarization in the RHIC range. The 2012 510 GeV inclusive jet \(A_{LL}\) extends gluon polarization measurements at lower \(x\) and agrees well with the 2009 200 GeV in the overlapping \(x_{T}\) range. STAR is continuing to produce important results in the near future and is planning upgrades for the further future that will probe gluon polarization down to \(x<10^{-3}\).


\end{document}